\documentclass[12pt, aps,showpacs,superscriptaddress,prl]{revtex4-1}
\usepackage{graphicx}
\usepackage{amsfonts,amssymb}
\usepackage{enumerate}
\usepackage[sort&compress]{natbib}
\usepackage{hyperref}
\usepackage[latin1]{inputenc}

\usepackage{amsfonts,amsmath,amssymb,graphicx,color,amsthm}
\usepackage{tgtermes}

\newtheorem*{thm}{Theorem}

\def\kk{{\bf k}}

\begin{document}

\title{Nonstationary distributions of wave intensities  in Wave Turbulence}

\author{Choi \surname{Yeontaek}}
\affiliation{National Institute for Mathematical Sciences, 70 Yuseong-daero 1689 beon-gil, Yuseung-gu, Daejeon 34047, Republic of Korea}
\author{Kwon \surname{Young-Sam}}
\affiliation{Department of Mathematics, Dong-A University, Busan 49315, Republic of Korea}

\author{Jo \surname{Sanggyu}}
\affiliation{Department of Physics, Kyungpook National University, Daegu 41566, Republic of Korea} 

\author{Nazarenko \surname{Sergey}} 
\affiliation{Mathematics Institute University of Warwick Coventry CV4 7AL UK}

\date{\today}

\begin{abstract}
 We obtain a general solution for the probability density function of wave intensities in non-stationary
Wave Turbulence. The solution is expressed in terms of the wave action spectrum evolving
according the the wave-kinetic equation. We establish that, in absence of wave breaking, the
wave statistics asymptotes to a Gaussian distribution in forced-dissipated wave systems that approach
a steady state. Also, in non-stationary systems,
if the statistics is Gaussian initially, it will remain Gaussian at any time.
Generally, the statistics that is not Gaussian initially will remain non-Gaussian over
the characteristic nonlinear time of the wave spectrum. In freely decaying wave turbulence, 
substantial deviations from Gaussianity may persist infinitely long.
\end{abstract}



\maketitle

\section{Introduction.} 
Wave Turbulence is a theory that describes random weakly nonlinear wave systems with broadband spectra (see e.g. ref.  \cite{naza11}). The main object in this theory is a wave action spectrum which is the second-order moment of the wave amplitude and which evolves according to the so-called wave-kinetic equation. Special attention in past literature was given to studies of stationary scaling solutions of this equation which are analogous to the Kolmogorov spectrum of hydrodynamic turbulence, the so-called Kolmogorov-Zakharov spectra. However, as it was shown in refs. \cite{lvov_2004,choi_2004,choi_2005,choi_2005b,naza11}, Wave Turbulence approach can also be extended to describing the higher-order moments and even to the entire probability density function (PDF) of the wave amplitude. A formal justification of such an extension based on a rigorous statistical
formulation was later presented in ref. \cite{Eyink}.
An introduction to Wave Turbulence as well as a summary recent developments in this area can be found in book  \cite{naza11}
and in an older text \cite{ZLF}.

It has been widely believed that the statistics of random weakly nonlinear wave systems
is close to being Gaussian. Derivation of the evolution equation for the PDF of the wave intensities presented in ref. \cite{choi_2005} has made it possible to examine  this belief.
It was shown  in ref. \cite{choi_2005} indeed has a stationary solution corresponding to the Gaussian state, but it was also noted that the typical evolution time of the PDF is the same as the one for the spectrum. Thus, for non-stationary wave systems one can expect significant deviations from the Gaussianity if the initial wave distribution is non-Gaussian. Note that non-Gaussian (typically deterministic) initial conditions for the wave intensity are typical in numerical simulations in Wave Turbulence. Also, there is no reason to believe that initial waves excited in natural conditions, e.g. sea waves excited by wind, should be Gaussian. Therefore, study of evolution of the wave statistics is important for both understanding of fundamental nonlinear processes as well as for the practical predictions such as e.g. wave weather forecast.

In the present paper we will present the full general solution for the PDF equation derived in  ref. \cite{choi_2005}. Based on that solution we will formulate the condition under which the wave statistics relaxes to the Gaussian state.

\section{Evolution equations for the wave amplitude and for the PDF}

Consider a weakly nonlinear wave system dominated by the four-wave interactions bounded by an  $L$-periodic cube in the $d$ dimensional physical space. (Four-wave systems are considered here as an illustrative example only. All results of this paper hold for the $N$-wave systems with any $N$. The only difference will be in the expressions for
$\gamma_\kk$ and  $\eta_\kk$ below; see ref. \cite{naza11}.) We have the Hamiltoninan equations for the Fourier coefficients as follows,
\begin{equation}
i\dot{a_{\bf k}} = \frac{\partial \mathcal{H}}{\partial {a}_{\bf k}^*}, \quad \mathcal{H}=\sum_{\bf k} \omega_{\bf k} |{a}_{\bf k}|^2 + \frac{1}{2}\sum_{{\bf k}_1, {\bf k}_2,{\bf k}_3,{\bf k}_4} W^{{\bf k}_1,{\bf k}_2}_{{\bf k}_3,{\bf k}_4}{a}_{{\bf k}_1}^* {a}_{{\bf k}_2}^* a_{{\bf k}_3} a_{{\bf k}_4} \label{4wH},
\end{equation}
where ${\bf k}, {\bf k}_1, {\bf k}_2,{\bf k}_3,{\bf k}_4 \in \frac L {2\pi} \mathbb{Z}^d $ are the wave vectors, $a_{\bf k}\in \mathbb{C}$ is the wave action variable, $W^{{\bf k}_1,{\bf k}_2}_{{\bf k}_3,{\bf k}_4} \in \mathbb{R}$ is an interaction coefficient which is a  model-specific function of 
${{\bf k}_1, {\bf k}_2,{\bf k}_3,{\bf k}_4}$
(e.g. $W^{{\bf k}_1,{\bf k}_2}_{{\bf k}_3,{\bf k}_4} =1$ for the Gross-Pitaevskii equation) and $\omega_{\bf k} \in \mathbb{R}$ is the frequency of mode $\bf k$.

Let us consider the PDF ${\mathcal P}(t,s_{\bf k})$
of the wave intensity $J_{\bf k}=
|a_{\bf k}|^2$ defined in a standard way as so that the probability for
$J_{\bf k}$ to be in the range from $s_{\bf k}$ to $s_{\bf k} +ds_{\bf k}$ is
${\mathcal P}(t,s_{\bf k}) d s_{\bf k}$. In symbolic form,
\begin{equation}
{\mathcal P}(t,s_{\bf k}) = \langle \delta (s_{\bf k}-J_{\bf k}) \rangle,\label{pasdelta}
\end{equation}

Suppose that the waves are weakly nonlinear, so that the quadric part of the Hamiltonian is much less than its quadratic part. Suppose also that the complex wave
amplitudes $a_{\bf k}$ are independent random variables for each  $\bf k$ and that the initial
phases of $a_{\bf k}$ are random and equally probable in the range from $0$ to $2\pi$.
These are the main assumptions of the weak Wave Turbulence theory (see ref.  \cite{naza11}), leading, upon taking the infinite-box limit $L \to \infty$, to the following evolution equation
 for  ${\mathcal P}(t,s_{\bf k})$, as derived in ref. \cite{choi_2005}:
\begin{equation}
\frac{\partial {\mathcal P}(t, s_{\bf k})}{\partial t} +\frac{\partial F}{\partial s_{\bf k}}=0,\label{main}
\end{equation}
where
\begin{equation}
F = -s_{\bf k} \Big(\gamma_{\bf k} {\mathcal P} +\eta_{\bf k} \frac{\partial {\mathcal P} }{\partial s_{\bf k}}\Big)
\end{equation}
and, for the four-wave systems,
\begin{eqnarray}
\eta_{\bf k}(t) &=& 4\pi \int|W^{{\bf k},{\bf k}_1}_{{\bf k}_2,{\bf k}_3}|^2\delta({{\bf k} + {\bf k}_1 -{\bf k}_2 -{\bf k}_2} )\delta(\omega_{{\bf k}} + \omega_{{\bf k}_1} - \omega_{{\bf k}_2}
-\omega_{{\bf k}_3}) n_{{\bf k}_1} n_{{\bf k}_2} n_{{\bf k}_3} d\kk_1d\kk_2d\kk_3, \\
\gamma_{\bf k}(t) &=& 8\pi \int|W^{{\bf k},{\bf k}_1}_{{\bf k}_2,{\bf k}_3}|^2\delta({{\bf k} + {\bf k}_1 -{\bf k}_2 -{\bf k}_2} )\delta(\omega_{{\bf k}} + \omega_{{\bf k}_1} - \omega_{{\bf k}_2}
-\omega_{{\bf k}_3}) \Big[n_{{\bf k}_1}(n_{{\bf k}_2}+n_{{\bf k}_3})-n_{{\bf k}_2}n_{{\bf k}_3}\Big]d\kk_1d\kk_2d\kk_3,
\label{gam}
\end{eqnarray}
where $n_{\bf k} = \langle J_{\bf k} \rangle$ is the wave action spectrum. The 
infinite-box limit resulted to passing continuous wave number description; each wave number integration in the above equations is over $\mathbb{R}^d$.

In this paper, we will find the time-dependent solution of the PDF  equation (\ref{main}). 

\section{Generating function }


%

Let us introduce the  generating function 
\begin{equation}
\mathcal{Z}(t,\lambda_{\bf k}) = \langle e^{-\lambda_{\bf k} |a_{\bf k}(t)|^2} \rangle =
\int^{\infty}_{0}{\mathcal P}(\lambda_{\bf k},t) e^{-\lambda_{\bf k} s_{\bf k}} ds_{\bf k}
\label{eq:Z}
\end{equation}
 where $\lambda_{\bf k}$ is a real parameter. Note that this definition is different from the one used in Ref.\cite{choi_2005} by the sign of the exponent. Here, we have changed the sign in order to comply with the standard relation between $\mathcal{P}$ and $\mathcal{Z}$ via the Laplace 
transform, as expressed in eqn. (\ref{eq:Z}).

In what follows we will drop subscripts $k$ for brevity whenever it does not cause ambiguity.

The  inverse Laplace transformation of  $\mathcal{Z}(t,\lambda)$ gives:
\begin{equation}
{\mathcal P}(t,s) =
\frac{1}{2\pi i}\lim_{T\to \infty} \int^{T+i\infty}_{T-i\infty}{\mathcal Z}(\lambda,t) e^{s\lambda} d\lambda.
\end{equation}
Given ${\mathcal Z}$, one can easily calculate the moments of the wave intensity,
\begin{equation}
M^{(p)}_{\bf k} \equiv \langle |a_{\bf k}|^{2p}\rangle =(-1)^p
{\mathcal Z}_{\lambda\cdots\lambda}|_{\lambda=0} = (-1)^p \langle |a|^{2p} e^{\lambda |a|^2}\rangle |_{\lambda=0},
\end{equation}
where $p \in \mathbb{N}$ is the order of the moments and subscript
$\lambda$ means taking derivative with respect to $\lambda$.
In particular, for the waveaction spectrum we have
$$
n_{\bf k} = - {\mathcal Z}_{\lambda}|_{\lambda=0} .
$$

The evolution equation for $\mathcal{Z}$ can be obtained by Laplace transforming
eqn. (\ref{eq:Z}), which gives
\begin{equation}
\dot{\mathcal{Z}} =- \lambda\eta \mathcal{Z}  -(\lambda^2\eta+\lambda\gamma)\mathcal{Z}_{\lambda}.\label{gf}
\end{equation}
Note that the sign differences in this equation with respect to the corresponding equation in Ref.\cite{choi_2005} is due to the sign difference in our definition of 
$\mathcal{Z}$.

Previously in Ref.\cite{choi_2005}, the general steady state solution eqn.(\ref{gf})
was presented:
\begin{equation}
{\mathcal{Z}} = \frac 1 {1+ \lambda_{\bf k} n_{\bf k}}.
\label{gfst}
\end{equation}
This solution corresponds to gaussian statistics of the wave field (Rayleigh distribution
for the wave intensity respectively).

Below, we will concentrate the fully time-evolving problem, in which the parameters $ \eta, \gamma$ have time-dependency. The goal of this paper is first to find the solution of the eqn. (\ref{gf}) and then obtain the respective time-dependent PDF. 

\section{Solution for $\mathcal{Z}$ by the method of characteristics }

We can find the solution for the fully time-evolving case of the eqn. (\ref{gf}) by using the  method of characteristics. Rewriting this equation in the characteristic form we have:

\begin{equation}
 \frac{d\lambda}{dt}= \Big({\gamma}+\lambda \eta \Big)\lambda, \quad \frac{d\mathcal{Z}}{dt}=- \lambda\eta\mathcal{Z}.\label{cm}
\end{equation}
  Changing  variable  to $\mu = \lambda e^{-\int_0^t {\gamma(t')}dt'}$
 in the first of these equations, we transform it
 into
\begin{equation}
\frac{d\mu(t)}{dt} =  \eta \mu \lambda =\eta \mu^2 e^{\int_0^t {\gamma(t')}dt'},
\label{cm1}
\end{equation}
solving which we have:
\begin{equation}
-\frac 1 {\mu(t)}+\frac 1 {\mu_0} =  \int^t_0 \eta(t') e^{\int^{t'}_0 \gamma(t'')dt''}dt',
\end{equation}

where $\mu_0 = \mu(0) = \lambda_0$. This gives for $\lambda(t) $:
\begin{equation}
\lambda(t) = \frac{\lambda_0 e^{\int^t_0 \gamma(t')dt'}}{1 - \lambda_0 \int^t_0 \eta(t') e^{\int^{t'}_0 \gamma(t'')dt''}dt'}.\label{slt}
\end{equation}
This relation has a simpler form in terms of $n$ rather than $\eta$. Indeed, $n$
satisfies the following (kinetic) equation,
\begin{equation}
\dot n = \eta - \gamma n,
\label{eq:ndot}
\end{equation}
integrating which we have
\begin{equation}
 n(t) =n(0) \, e^{-\int^t_0 \gamma(t')dt'} + e^{-\int^t_0 \gamma(t')dt' } \int^t_0 \eta(t') e^{\int^{t'}_0 \gamma(t'')dt''}dt'.
\label{eq:n}
\end{equation}
Using this identity, we have:
\begin{equation}
\lambda(t) = \frac{\lambda_0 }{e^{-\int^t_0 \gamma(t')dt'} - \lambda_0 \left(n(t)-n(0) e^{-\int^t_0 \gamma(t')dt'}\right)}.\label{slt1}
\end{equation}
Conversely,
 we have:
\begin{equation}
\lambda_0 = \frac{\lambda e^{-\int^t_0 \gamma(t')dt'}  }{1+ \lambda \left(n(t)-n(0) e^{-\int^t_0 \gamma(t')dt'}\right)}.\label{slt1i}
\end{equation}

Now, from the second of the eqns. (\ref{cm}) and from the first equality in
eqn. (\ref{cm1}) we see that the log derivative of $\mathcal{Z}$ is equal
to the negative log derivative of $\mu$. Thus,
\begin{equation}
\mathcal{Z}(t, \lambda)=\mathcal{Z}_0 \frac  {\mu_0} \mu=
\frac   {\mathcal{Z}_0 \lambda_0}  \lambda e^{\int_0^t {\gamma(t')}dt'}
=
\frac  { \mathcal{Z}_0 }{1+ \lambda \left(n(t)-n(0) e^{-\int^t_0 \gamma(t')dt'}\right)}.\label{solnZ}
\end{equation}
where $\mathcal{Z}_0 = \mathcal{Z}(0, \lambda_0)$ and 
$\lambda_0$ must be substituted in terms of $\lambda$ solving for it from
eqn. (\ref{slt1i}).

Eqns. (\ref{solnZ}) and (\ref{slt1i}) allow us to prove the following theorem.

\begin{thm} 
\begin{enumerate}
\item Wave fields which are Gaussian initially will remain Gaussian for all time.
\item Wave turbulence asymptotically becomes Gaussian if
  \begin{equation}
\lim_{t \to \infty} \frac {n(0) e^{-\int_0^t {\gamma(t')}dt' }}{n(t)} = 0.
\label{cond}
\end{equation}
\end{enumerate}
\end{thm}

To prove the first part we simply substitute 
$\mathcal{Z}_0  = 1/(1+ \lambda_0 n_0)$ into eqn. (\ref{solnZ})
and, after using  (\ref{slt1i}), obtain $\mathcal{Z} = 1/(1+ \lambda n)$,
which corresponds to the Gaussian statistics.

To prove the second part we notice that if condition  (\ref{cond}) is satisfied then
  \begin{equation}
\lim_{t \to \infty} \lambda_0 (t, \lambda) = 0, \quad
\lim_{t \to \infty} \mathcal{Z}_0 = \mathcal{Z}(0, 0) =1
\quad \hbox{and} \quad
\lim_{t \to \infty} \mathcal{Z}(t, \lambda) =
\frac  { 1 }{1+ \lambda n}.
\label{cond1}
\end{equation}

{\bf Remarks:}
\begin{enumerate}
\item
Condition (\ref{cond}) is satisfied for the inertial range modes in forced-dissipated systems which tend to a steady state. Indeed, in this case $\gamma \to \eta/n $ which is a
positive constant (at fixed mode $\bf k$), so the time integral of this quantity diverges as $t \to \infty$.
\item
In absence of forcing and dissipation, spectrum $n_{\bf k}$ decays to zero at
any mode $\bf k$ as $t \to \infty$, and so does $\gamma_{\bf k}$. Thus the integral
of $\gamma_{\bf k}(t)$ may converge as $t \to \infty$, which means that non-Gaussianity of some (or all)
wave modes may persist  as  $t \to \infty$.
\item
In general, function  $\gamma_{\bf k}(t)$ is not sign definite, and there may be transient
time periods where  $\gamma_{\bf k}(t)<0$. The deviation from Gaussianity of some (or all)
wave modes may increase during these periods.
\end{enumerate}

\section{Evolution of the PDF}

Now let us analyse the PDF of transient states.
Let us think of a simple case with a deterministic initial wave intensity, $P(0,s) = \delta(s - J)$. We will call such a  solution $\mathcal{P}_J(s,t)$. Then $\mathcal{Z}(0,\lambda) = e^{-\lambda J}$. 
In fact, since the inverse Laplace transform is a linear operation, the considered solution is nothing but Green's function for the general problem with an arbitrary initial condition $P(0,s)$:
\begin{equation}
\mathcal{P}(t,s) =\int_0^\infty \mathcal{P}(0,J) \mathcal{P}_J(t,s) dJ. \label{solnPgen}
\end{equation}

Let us take the inverse Laplace transform of  $\mathcal{Z}(t, \lambda)$ given by
eqn. (\ref{solnZ}) to obtain the $\mathcal{P}_\delta(s,t) $ at $t>0$:
\begin{equation}
\mathcal{P}_J(t,s) = \frac{1}{2\pi i}\lim_{T\to +\infty}  \int^{T+i\infty}_{T-i\infty} e^{\lambda s}\mathcal{Z}(\lambda) d\lambda \\ 
=  \frac{1}{2\pi i} \lim_{T\to +\infty}  \int^{T+i\infty}_{T-i\infty} 
\frac  { e^{\lambda s-\lambda_0 J}  }{1+ \lambda \tilde n }
 d\lambda.  \label{solnPst}
\end{equation}
where 
\begin{equation}
\tilde n = n(t)-J e^{-\int^t_0 \gamma(t')dt'}
\end{equation}
(note that $n(0) =J$).
Substituting $\lambda_0$ from (\ref{slt1}) and changing the integration
variable as $ \rho = \lambda +1/\tilde n$,
we have:
\begin{equation}
\mathcal{P}_J(t,s) 
= \frac{e^{-\frac{s}{\tilde n} - a \tilde n}}{2\pi i {\tilde n}}
\lim_{T\to +\infty} \int^{T+i\infty}_{T-i\infty} 
\frac{ e^{s \rho +\frac a \rho} }\rho d \rho
=\frac 1 { {\tilde n}} {e^{-\frac{s}{\tilde n} - a \tilde n}}
I_0(2\sqrt{as})
.  \label{solnPst1}
\end{equation}
where $a = \frac{J}{ {\tilde n}^2} e^{-\int^t_0 \gamma(t')dt'}$
 and $I_0(x) $ is the zeroth modified Bessel function of the first kind.
Note that $I_0(0)=1$, so we recover $\mathcal{P}_\delta \to \mathcal{P}_G = \frac 1 n e^{-s/n}$
as $t \to \infty$ if condition (\ref{cond}) is satisfied provided that $s$ is not too large, $as \ll 1$.

Now let us suppose that condition (\ref{cond}) is satisfied and
let consider the asymptotic behaviour of the probability distribution at large $s$ 
and large $t$, and $as \gg 1$ (i.e.  $s$ is much larger than $1/a$ which is itself large). 
Taking into account that $I_0(x) \xrightarrow{x\rightarrow \infty} \frac{e^x}{\sqrt{2\pi x}}$, we have:
\begin{equation}
\mathcal{P}_J(s,t) \to 
\frac{\mathcal{P}_G}{(2\pi)^{1/2}(as)^{1/4}}e^{2\sqrt{as} - as}  \ll
\mathcal{P}_G \quad \hbox{for} \quad   as\gg 1, \; \int^t_0 \gamma(t')dt' \gg 1.
\label{front}
\end{equation}
Thus, we see a front at $s \sim s^{*}(t) = 1/a $ moving toward large $s$ as $t \to \infty$. The PDF ahead of this front is depleted
with respect to the Gaussian distribution, whereas behind the front it asymptotes
to  Gaussian. Obviously, the same kind of behaviour will be realised for any solution
(\ref{solnPgen}) arising from initial data having a finite support in $s$.

\section{Conclusions and Discussion}

In this paper we have obtained the general solutions for the 
generating function and for the PDF of wave intensities
in Wave Turbulence, equations (\ref{solnZ}) and (\ref{slt1i}),
and equation (\ref{solnPst1}) respectively.
This allowed us to prove a theorem stating
that wave fields which are Gaussian initially will remain Gaussian for all time
and that
 Wave Turbulence asymptotically becomes Gaussian if condition (\ref{cond})
is satisfied. We have also found (when  condition (\ref{cond})
is satisfied) an asymptotic solution for the PDF (\ref{front}) where
 the Gaussian distribution forms behind a front propagating toward large wave intensities.

Condition (\ref{cond}) is satisfied for the inertial range modes in forced-dissipated systems approaching a steady state. 
Thus, the Gaussian statistics will form at large time for such modes in these systems.
An interesting subclass of solutions in forced-dissipated systems is when the
spectrum is in a steady state from the initial moment of time (i.e. it is a stationary solution
of the wave-kinetic equation), while the 
PDF is not Rayleigh initially (i.e. the initial wave field is not Gaussian).
For example, the initial wave intensities can be deterministic, i.e. their PDFs are delta-functions,
as it is often taken in numerical simulations of Wave Turbulence.
In this case, equation (\ref{solnPst1}) looks the simplest, with
$\int^t_0 \gamma(t')dt' = \gamma t$.

Since the characteristic evolution times are the same for the
spectrum $n_{\bf k}$ and the PDF, the latter will remain non-Gaussian
over a substantial time in the initial field is non-Gaussian.
Such situations should be considered typical rather than exception in
natural conditions (where initial waves arise, e.g., from an instability
which does not necessarily produce Gaussian waves) and in numerical simulations 
(where typically the wave intensities are taken to be deterministic).

Moreover, in absence of  forcing and dissipation, spectrum $n_{\bf k}$ decays to zero at
any mode $\bf k$ as $t \to \infty$, and so does $\gamma_{\bf k}$. Thus the integral
 $\int^t_0 \gamma_\kk(t')dt' $ may converge as $t \to \infty$, which means that non-Gaussianity of some (or all)
wave modes may persist  as  $t \to \infty$.
Furthermore, since   $\gamma_{\bf k}(t)$ is not sign definite,  there may be transient
time periods where  $\gamma_{\bf k}(t)<0$. The deviation from Gaussianity of some (or all)
wave modes may increase during these periods.

The present paper considers the four-wave systems as an illustrative example,
but it is clear that the obtained solutions are more general and apply to the wave systems
with resonances of any order (one would simply have to use different expressions
for the integrals $\gamma_\kk(t)$ and $\eta_\kk(t)$ 
corresponding to resonance of the considered order; see e.g. book \cite{naza11}).
Note that our solution for the PDF (\ref{solnPst1}) is expressed in terms of the spectrum
$n_\kk(t)$  (recall that $\gamma_\kk(t)$  depends on $n_\kk(t)$
via eqn. (\ref{gam})). On the other hand, $n_\kk(t)$ obeys the wave-kinetic
equation which is not easy to solve for non-stationary systems.
However, it is quite straightforward to solve the wave-kinetic
equation numerically, after which the resulting $n_\kk(t)$ can be used
in the analytical formula  for the PDF (\ref{solnPst1}).
Since the latter formula is very simple, we believe that it can
be very effective in practical calculations
especially in the situations where non-Gaussianity is important, e.g., in
wave weather forecasts including prediction of anomalously strong waves --
the so-called freak waves.

\begin{acknowledgments}

Yeontaek Choi's work is supported by National Institute for Mathematical Sciences(NIMS) and Korean Union of Public sector Research and Professional workers(KUPRP). The work of Young-Sam Kwon is supported by Basic Science Research Program through the National Research Foundation of Korea(NRF) funded by the Ministry of Education(NRF-2013R1A1A2057662).

\end{acknowledgments}


%


\end{document}